\documentclass[a4paper]{jpconf}

\usepackage{graphicx}
\begin{document}
\title{Ultra High Energy Cosmic Rays and the Highest Energies Universe}

\author{Roberto Aloisio}

\address{Gran Sasso Science Institute, L'Aquila, Italy.\\
INFN - Laboratori Nazionali Gran Sasso, Assergi (AQ), Italy.}

\ead{roberto.aloisio@gssi.it}

\begin{abstract}
We will discuss the main relevant aspects of the physics of ultra high energy cosmic rays. After a short recap of the experimental evidences, we will review theoretical models aiming at describing the sources of these extremely energetic particles opening a window on the highest energies universe. We will discuss the production of secondary particles and the possible tests of new physics that ultra high energy cosmic rays could provide. The present proceedings paper is mainly based on the review papers \cite{Aloisio:2017ooo,Aloisio:2017qoo}.
\end{abstract}

\section{Introduction}
The study of the highest energies universe is inextricably tied to Ultra High Energy Cosmic Rays (UHECR) the highest energy particles ever observed with energies ranging from $10^{17}$ eV up to energies in excess $10^{20}$ eV. The observation of UHECR brings informations about the most energetic events in the universe and it could unveil new physical phenomena at energetic regimes not accessible otherwise.

The most advanced experiments to detect UHECR are, nowadays, the Pierre Auger Observatory in Argentina \cite{ThePierreAuger:2015rma} and the Telescope Array (TA) experiment in the US \cite{Tinyakov:2014lla}, covering roughly 1/10 of the Auger area. The observation of UHECR is performed through their interaction with the Earth atmosphere, that produces an Extensive Air Shower (EAS) of particles in the atmosphere with a correlated fluorescence emission due to the excitation/de-excitation of nitrogen molecules in the atmosphere during the EAS development (see \cite{Aloisio:2017ooo} and references therein). 

Both Auger and TA are based on the hybrid concept, they combine the observation of the EAS particles that reach the ground and the observation of the fluorescence emission produced by the EAS in the atmosphere. This is obtained combining an array of surface detectors (SD) to sample EAS particles when they reach the ground and telescopes, overlooking the surface array, to collect the fluorescence light (fluorescence detectors, FD). The possibility of performing a combined measurement of SD and FD was one the most important breakthrough in the field of UHECR, this combination enables a better determination of the energy scale through the FD and the collection of high statics through the SD. 

The experimental study of UHECR clarified few important facts: (i) UHECR are charged particles, with a limit on photon and neutrino fluxes at $10^{19}$ eV at the level of few percent and well below respectively \cite{Abraham:2009qb,Abu-Zayyad:2013dii,Abreu:2013zbq}, (ii) the spectrum observed at the Earth shows a slight flattening at energies around $5\times 10^{18}$ eV (called the ankle) with (iii) a steep suppression at the highest energies \cite{ThePierreAuger:2013eja,Abu-Zayyad:2013qwa}.

The mass composition of UHECR is still not completely determined with experimental evidences showing a certain level of discrepancy. Before Auger published its results, experimental evidences were all pointing toward a prevalently light composition (proton dominated) with acceleration models characterised by steep injection spectra and high maximum acceleration energies (up to energies larger than $10^{20}$ eV) \cite{Aloisio:2017ooo,Aloisio:2017qoo,Berezinsky:2002nc}.

The observations of Auger, far the largest experiment set-up devoted to the detection of UHECR, have shown that the UHECR mass composition is dominated by protons only at energies around and below $10^{18}$ eV becoming progressively heavier starting from $5\times 10^{18}$ eV, with the highest energies dominated by nuclei \cite{Aab:2014kda}. On the other hand, the TA experiment, even if with $1/10$ of the Auger statistics, collected data that seem to confirm the pre-Auger scenario \cite{Abbasi:2014sfa}, the mass composition is compatible with being light for energies above $10^{18}$ eV, with no apparent transition to a heavier mass composition. Nevertheless, a joint working group made of members of both collaborations, TA and Auger, has recently concluded that the results of the two experiments are not in conflict once systematic and statistical uncertainties have been taken into account \cite{Abbasi:2015xga}. This conclusion, though encouraging on one hand, casts serious doubts on the possibility of reliably measuring the mass composition at the highest energies, unless some substantially new piece of information becomes available. 

It should be also noted that the spectra measured by the two experiments, though being in general agreement, differ beyond the systematic error at the highest energies (where mass composition differs the most) in such a way that TA claims a spectral suppression at energy $\ge 5\times 10^{19}$ eV while Auger shows the suppression at sensibly lower energies \cite{Verzi:2015dna}.

In order to interpret the observations at the Earth, it is important a detailed modelling of UHECR propagation in the intergalactic medium, which is mainly conditioned by the interaction with astrophysical photon backgrounds. These interactions shape the spectrum observed at the Earth and are also responsible for the production of secondary (cosmogenic) particles: photons and neutrinos. This secondary radiation can be observed through ground-based or satellite experiments and brings important informations about the mass composition of UHECR and, possibly, on their sources. 

Sources of UHECR are still a mystery, we do not know which kind of astrophysical object is responsible for the production of these particles. There are basically two different classes of astrophysical mechanisms that could be invoked to accelerate UHECR \cite{Aloisio:2017ooo,Aloisio:2017qoo}. The first class is based on the transfer of energy from a macroscopic object (that can move relativistically or not) through repeated interactions of particles with magnetic inhomogeneities; belongs to this class the diffusive shock acceleration mechanism. The second class is based on the interaction with electric fields that, through high voltage drops, can accelerate particles (at  once) until the highest energies; belongs to this class the unipolar induction mechanism in fast spinning neutron stars. At the highest energies $E>5\times 10^{19}$ eV, the production of UHECR could also be connected with new physics not testable in Earth's laboratories. This is the case of the so-called top-down models in which UHECR are directly produced at high energy, as decay products of super-heavy relic particles (with mass $M_X \simeq 10^{12}\div 10^{14}$ GeV) predicted in a wide class of inflationary scenarios. These models connect the observations of UHECR with the Dark Matter (DM) problem and cosmological observations \cite{Aloisio:2015lva}.

The paper is organised as follows. In section \ref{sec:models} we will discuss the main features of the acceleration models, compatible with observations, and their implications in terms of possible astrophysical sources. In section \ref{sec:cosmogenic} we will discuss the production of secondary (cosmogenic) particles during the propagation of UHECR, mainly focusing on secondary neutrinos. In section \ref{sec:SHDM} we will discuss the impact on the physics of UHECR of exotic models aimed at solving the DM problem. We will conclude in section \ref{sec:concl}.

\section{Theoretical models and possible sources}
\label{sec:models}
To constrain the basic features of the sources of UHECR it is useful to adopt a purely phenomenological approach in which sources are homogeneously and isotropically distributed, characterised by few parameters: the injection of particles by the sources with a power law behaviour in energy $\propto E^{-\gamma_g}$, the maximum acceleration energy $E_{max}$ and the source emissivity ${\cal L}_S(A)$ , i.e. number of particles injected per unit volume and energy for each nuclei species labeled by the atomic mass number $A$. These parameters, that univocally identify a class of sources, are fitted to experimental data both spectrum and mass composition (see \cite{Aloisio:2017ooo,Aloisio:2017qoo} and references therein). 

Following \cite{Aloisio:2013hya,Aloisio:2015ega}, taking into account all possible channels of energy losses and solving the transport equations for UHECR (protons or nuclei) we can determine the theoretical flux and mass composition expected at the Earth, to be compared with observations. While the flux directly follows solving the transport equations, mass composition is inferred from the mean value of the depth of shower maximum (elongation rate) $\langle X_{max} \rangle$ and its dispersion $\sigma(X_{max})$, computed as shown in \cite{Abreu:2013env}. The combined analysis of $\langle X_{max} \rangle$ and $\sigma(X_{max})$, even if not conclusive allows to obtain less model dependent information on the mass composition \cite{Aloisio:2007rc,Kampert:2012mx}. A relevant source of uncertainties comes from the dependence of the depth of shower maximum and its fluctuations on the hadronic interaction model used to describe the shower development in the atmosphere. Most of such models fit low energy accelerator data while providing somewhat different results when extrapolated to the energies of relevance for UHECR (see \cite{Aloisio:2017ooo,Aloisio:2017qoo} and references therein).

As discussed in the Introduction, the results of Auger showed a mass composition progressively dominated by heavy nuclei. The qualitative new finding that mass composition might be mixed has served as a stimulus to build models that can potentially explain the phenomenology of Auger data. These models all show that the Auger spectrum and mass composition at $E\ge 5\times 10^{18}$ eV can be fitted at the same time only at the price of requiring very hard injection spectra for all nuclei and a maximum acceleration energy $E_{max}\le 5 Z\times 10^{18}$ eV \cite{Aloisio:2013hya,Aloisio:2009sj}, being $Z$ the charge of the nucleus. The need for hard spectra can be understood taking into account that the low energy tail of the flux of UHECR reproduces the injection power law. Therefore, taking $\gamma\ge 2$ cause the low energy part of the spectrum to be polluted by heavy nuclei thereby producing a disagreement with the light composition observed at low energy.

One should appreciate the change of paradigm that these findings imply: while in the case of a pure proton composition it is needed to find sources and acceleration mechanisms able to energise CR protons up to energies larger than $10^{20}$ eV with steep injection \cite{Aloisio:2006wv} ($\gamma_g\simeq 2.5\div 2.7$), the Auger data require that the highest energy part of the spectrum ($E>5\times 10^{18}$ eV) has a flat injection ($\gamma_g\simeq 1.0\div 1.6$) being dominated by heavy nuclei with maximum energy not exceeding a few$\times Z\times 10^{18}$ eV  \cite{Aloisio:2013hya}. 


\begin{figure}[!h]
\centering
\includegraphics[scale=.28]{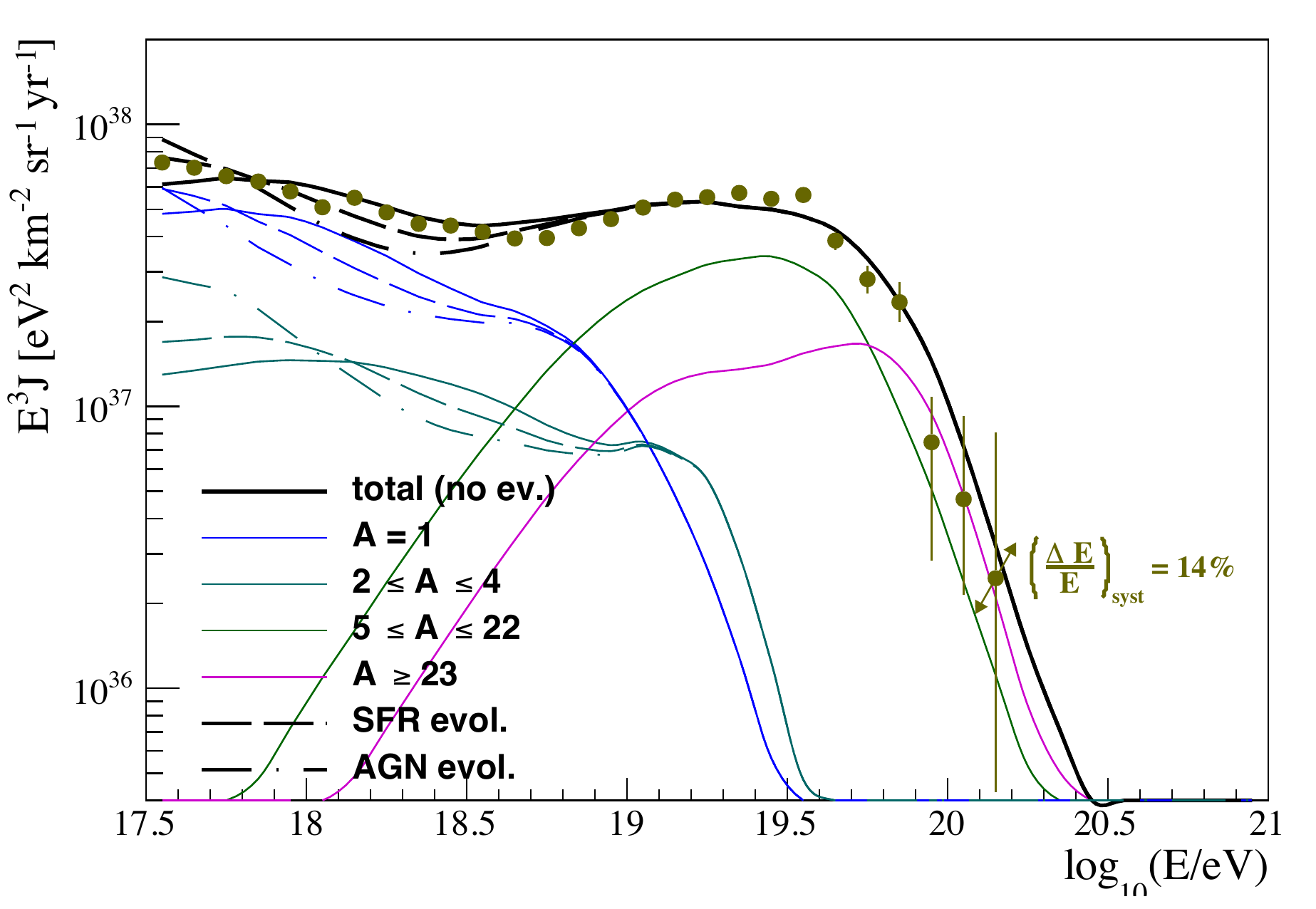}
\includegraphics[scale=.28]{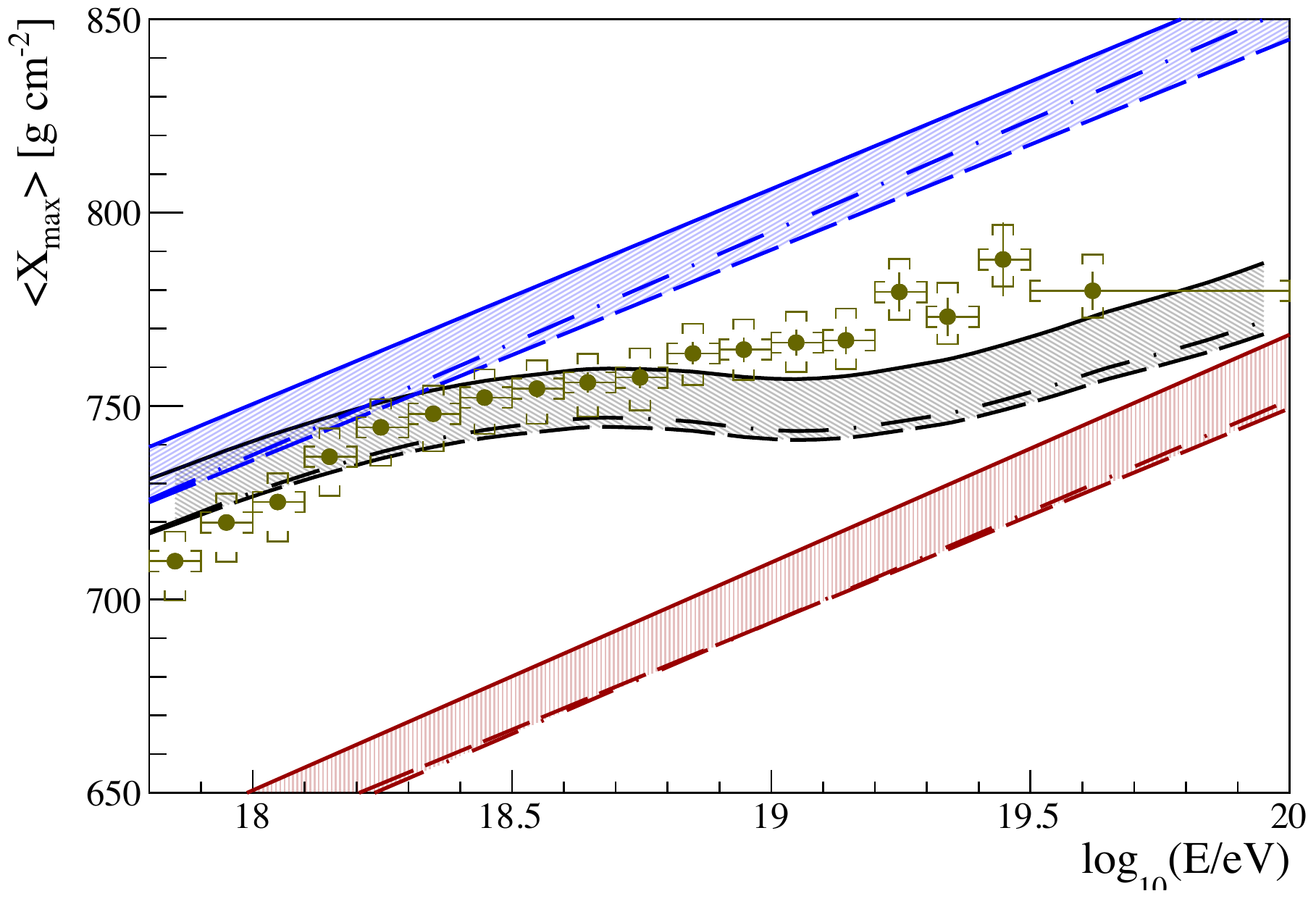}
\includegraphics[scale=.28]{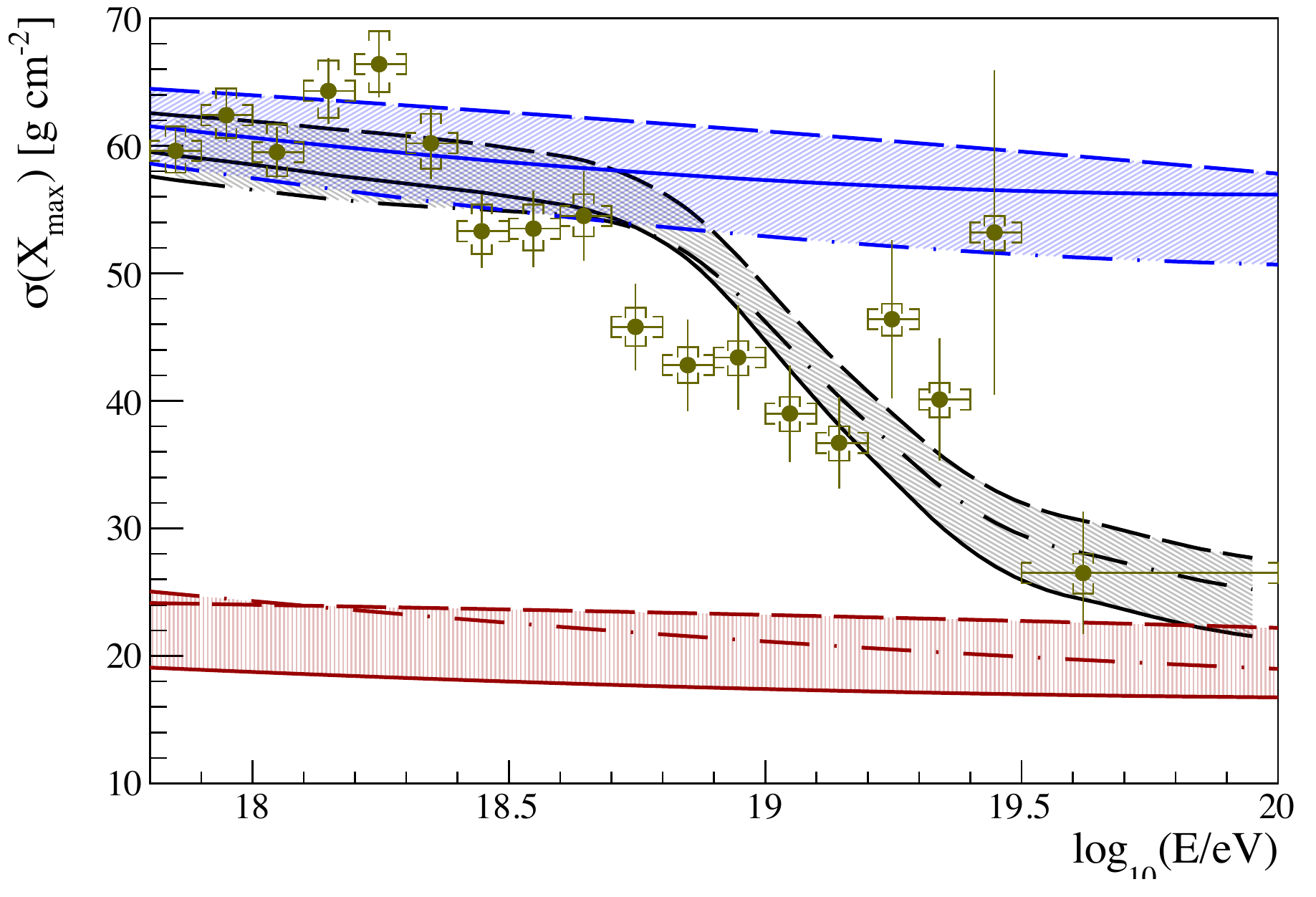}
\caption{Comparison of the flux (left panel), elongation rate (central panel) and its dispersion (right panel) as observed by Auger and computed assuming the model with two classes of different extragalactic sources. Figures taken from \cite{Aloisio:2015ega}.}
\label{fig1}  
\end{figure}

In order to reproduce Auger observations, the additional light contribution to the flux at energies below $5\times 10^{18}$ eV should exhibit a steep power law injection with $\gamma_g\simeq 2.6\div 2.7$ and a maximum acceleration energy not exceeding a few$\times 10^{18}$ eV, as for the heavier component \cite{Aloisio:2013hya,Taylor:2013gga,Globus:2015xga}. The origin of this radiation that, as for the heavier one, should be extragalactic \cite{Aloisio:2017ooo,Aloisio:2017qoo}, can be modelled essentially in two ways: (i) assuming the presence of different classes of sources: one injecting also heavy nuclei with hard spectrum and the other only proton and helium nuclei with soft spectrum \cite{Aloisio:2013hya,Taylor:2013gga} or (ii) identifying a peculiar class of sources that could provide at the same time a steep light component and a flat heavy one \cite{Globus:2015xga,Unger:2015laa,Blasi:2015esa}. The second approach is based on a specific hypothesis on the sources that should be surrounded by an intense radiation field that, through photo-disintegration of heavy nuclei in the source neighbourhood, can provide a light component of (secondary) protons with a steep spectrum together with a hard and heavier component \cite{Globus:2015xga,Unger:2015laa}. It should be also noted that the diffusive shock acceleration mechanism produces steep injections of particles with $\gamma_g > 2$, while acceleration due to potential drops typically imply $\gamma_g\simeq 1$ (see \cite{Aloisio:2017ooo,Aloisio:2017qoo} and references therein). 

In figure \ref{fig1} we plot the flux and mass composition as observed by Auger and as reproduced theoretically assuming two classes of extragalactic sources with different injection characteristics as discussed above.

\section{Cosmogenic secondary neutrinos} 
\label{sec:cosmogenic}

The propagation of UHECR is mainly conditioned by the interaction with astrophysical photon backgrounds: the Cosmic Microwave Background (CMB) and the Extragalactic Background Light (EBL). The interaction of UHECR with such backgrounds shapes the spectrum expected at the Earth and produces several unstable particles that in turn decay into high energy photons, electrons/positrons and neutrinos. The possible detection of these signal carriers, realised already at the end of sixties, is extremely important to constrain models for UHECR sources, composition and the details of propagation (see \cite{Aloisio:2017ooo,Aloisio:2017qoo} and references therein). In the following we will mainly discuss the case of neutrinos.

There are two processes by which neutrinos can be emitted in the propagation of UHECR: (i) the decay of charged pions, produced by photo-pion production, $\pi^{\pm}\to \mu^{\pm}+\nu_{\mu}(\bar{\nu}_{\mu})$ and the subsequent muon decay $\mu^{\pm}\to e^{\pm}+\bar{\nu}_{\mu}(\nu_{\mu})+\nu_e(\bar{\nu}_e)$; (ii) the beta-decay of neutrons and nuclei produced by photo-disintegration: $n\to p+e^{-}+\bar{\nu}_e$, $(A,Z)\to (A,Z-1)+e^{+}+\nu_e$, or $(A,Z)\to (A,Z+1)+e^{-}+\bar{\nu}_e$. These processes produce neutrinos in different energy ranges: in the former the energy of each neutrino is around a few percent of that of the parent nucleon, whereas in the latter it is less than one part per thousand (in the case of neutron decay, larger for certain unstable nuclei). This means that in the interaction with CMB photons, which has a threshold Lorentz factor of about $\Gamma\ge 10^{10}$, neutrinos are produced with energies of the order of $10^{18}$~eV and $10^{16}$~eV respectively. Interactions with EBL photons contribute, with a lower probability than CMB photons, to the production of neutrinos with energies of the order of $10^{15}$~eV in the case of photo-pion production and $10^{14}$~eV in the case of neutron decay (see \cite{Aloisio:2015ega} and reference therein).   

\begin{figure}[!h]
\centering
\includegraphics[scale=.43]{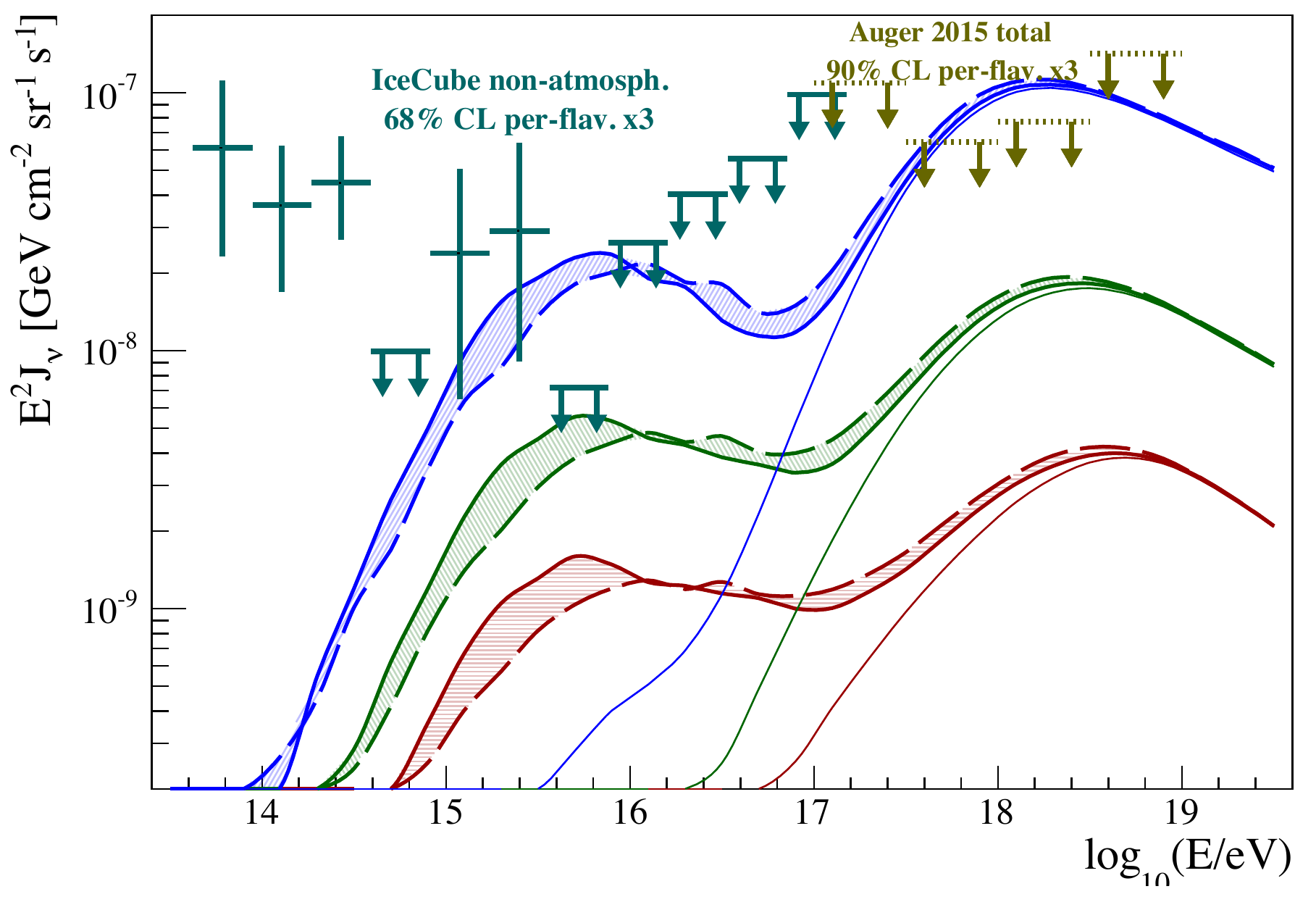}
\includegraphics[scale=.43]{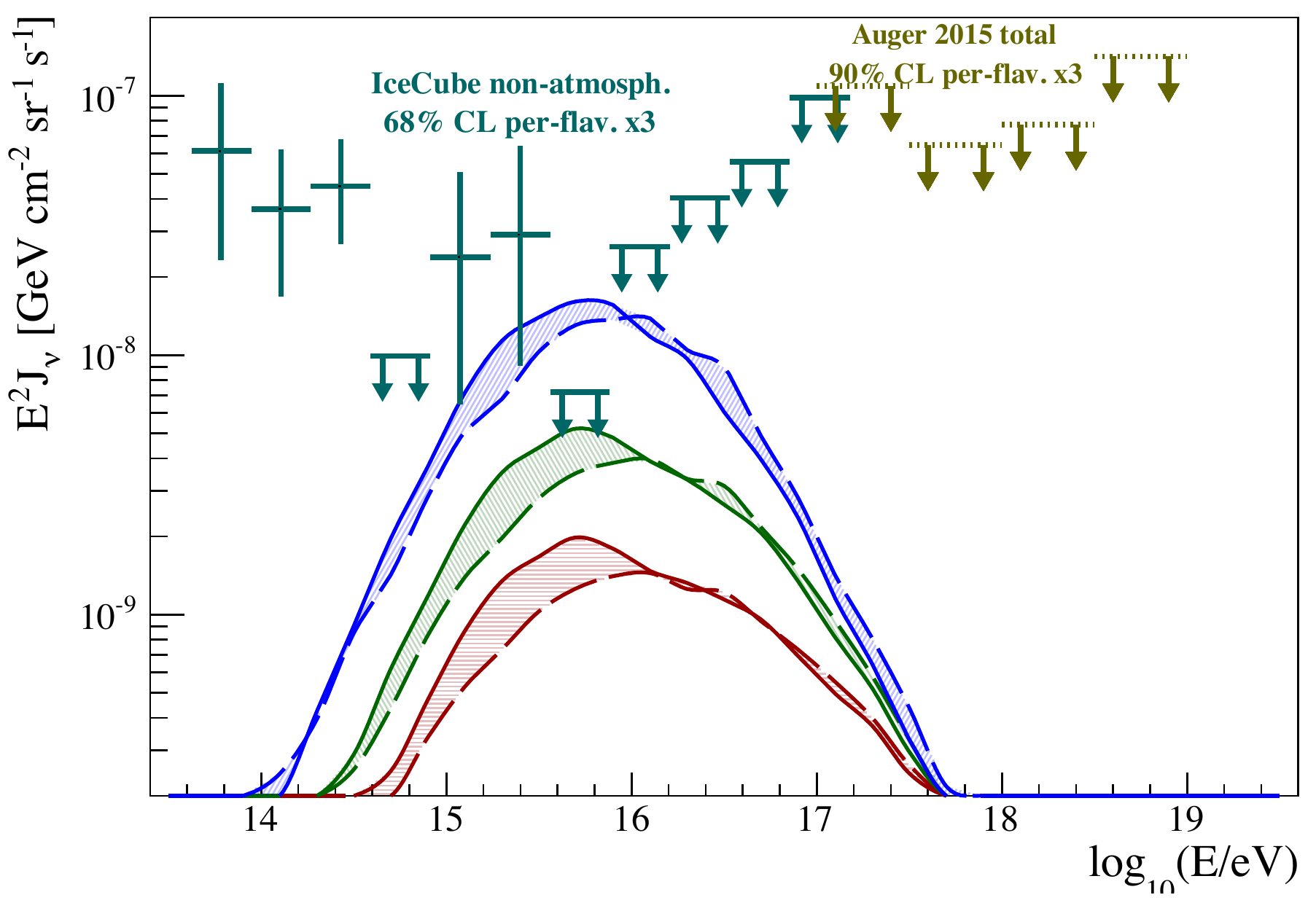}
\caption{ [Left panel] Fluxes of neutrinos in the case of a pure proton composition. The three different fluxes correspond to different assumptions on the cosmological evolution of sources (from bottom to top): no evolution (red), star formation rate evolution (green) and active galactic nuclei evolution (blue), coloured bands show the uncertainties due to the EBL model (see \cite{Aloisio:2017qoo,Aloisio:2017ooo} and references therein). Thin solid lines are neutrino fluxes obtained taking into account the sole CMB field. [Right Panel] Neutrino fluxes in the case of mixed composition, as shown in figure \ref{fig1}, with the same color code of left panel. Experimental points are the observations of IceCube on extra-terrestrial neutrinos \cite{Aartsen:2015awa} and the Auger limits on neutrino fluxes \cite{Abreu:2013zbq}. Figures taken from \cite{Aloisio:2015ega}.} 
\label{fig2}
\end{figure}

The flux of secondary neutrinos is very much sensitive to the composition of UHECR. In figure \ref{fig2} we plot the flux of cosmogenic neutrinos expected in the case of a pure proton composition (left panel) and in the case of mixed composition (right panel). Comparing the two panels of figure \ref{fig2} it is evident the huge impact of the composition on the expected neutrino flux: heavy nuclei provide a reduced flux of neutrinos because the photo-pion production process in this case is subdominant. 

The production of cosmogenic neutrinos is almost independent of the variations in sources' distribution because the overall universe, up to the red-shift of the first stars (pop III) $z_{max}\simeq 10$, could contribute to the flux. Once produced at these cosmological distances neutrinos travel toward the observer almost freely, except for the adiabatic energy losses and flavour oscillations, the opacity of the universe to neutrinos being relevant only at redshifts $z \gg 10$. This is an important point that makes neutrinos a viable probe not only of the mass composition of UHECR but also of the cosmological evolution of sources. In figure \ref{fig2} three different hypothesis on the cosmological evolution of sources are taken into account: no cosmological evolution (red bands), evolution typical of the star formation rate (green band) and of active galactic nuclei (blue band). 
 
There is a solid consensus about the light composition of UHECR in the low energy part of the observed spectrum. This assures a flux of cosmogenic neutrinos in the PeV energy region, produced by the protons' photo-pion production on the EBL photons. Coloured bands in figure \ref{fig2} show the uncertainties connected with the EBL background (see \cite{Aloisio:2017ooo,Aloisio:2017qoo} and references therein). Another important uncertainty in the expected neutrino flux comes from the contribution of UHECR sources at high red-shift. Given the energy losses suffered by UHE protons and nuclei, sources at red-shift larger than $z>1$ can be observed only in terms of cosmogenic neutrinos \cite{Aloisio:2015ega}. 

\begin{figure}[!h]
\centering
\includegraphics[scale=0.62]{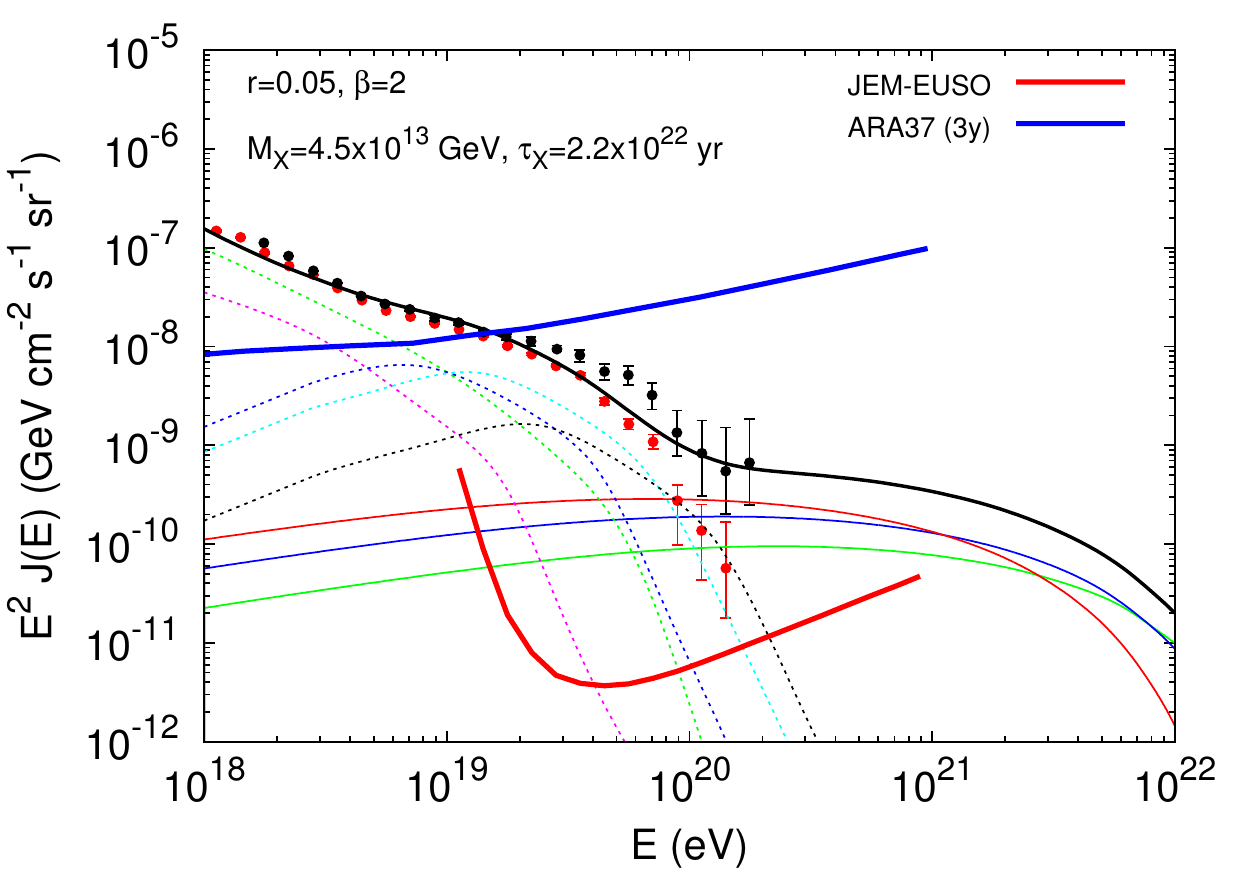}
\includegraphics[scale=0.62]{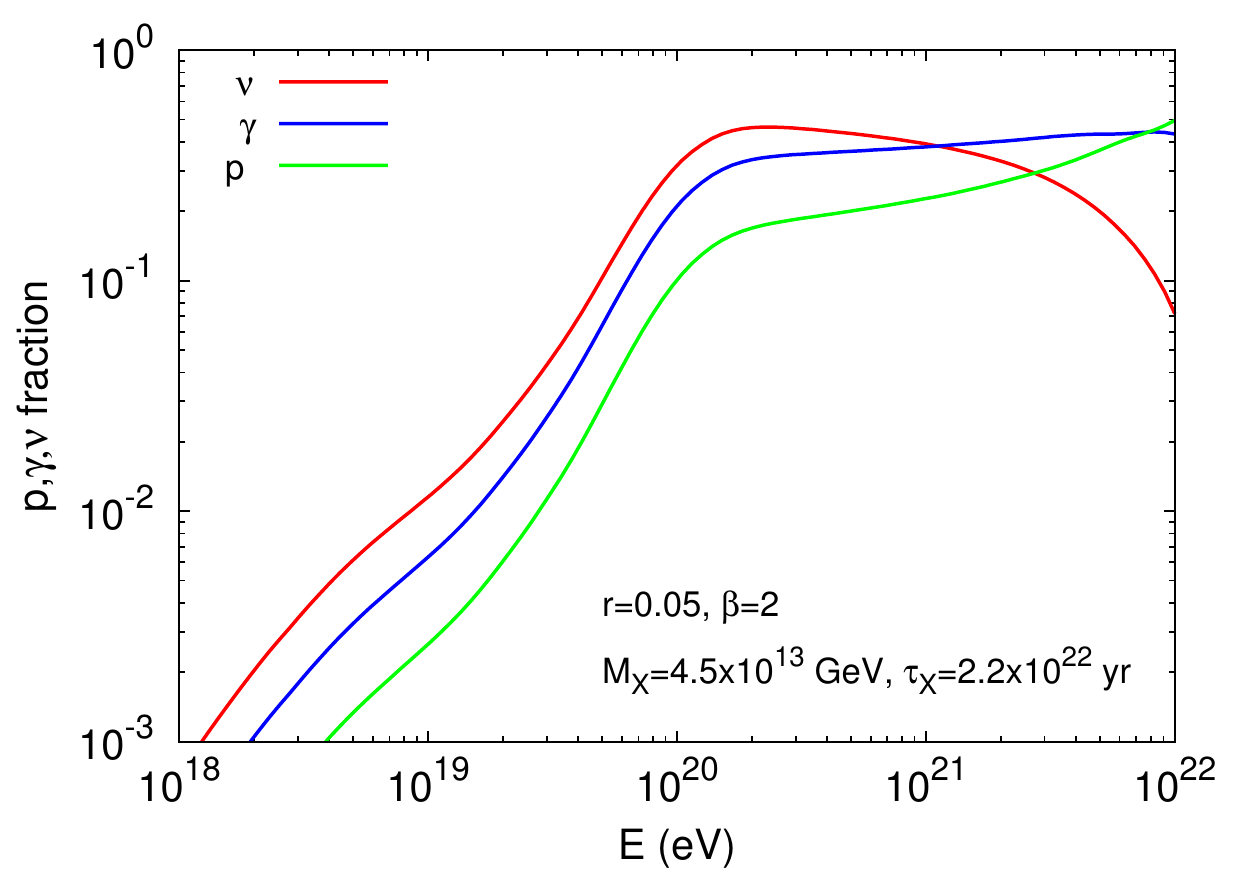}
\caption{[Left Panel] Flux of UHECR from the decay of SHDM (thin solid lines) with parameters as labeled together with the flux expected in the framework of the mixed composition model of \cite{Aloisio:2015ega}. Also shown is the sensitivity to SHDM decay products: of the proposed JEM-EUSO space mission (thick red solid line) and, for UHE neutrinos, the upcoming ARA observatory (thick blue solid line). Experimental data are those of Auger (red points) and TA (black points). [Right Panel] Fraction over the total flux of protons, photons and neutrinos by SHDM decay as follows from fluxes in right panel. Figures taken from \cite{Aloisio:2015lva}.} 
\label{fig3}   
\end{figure}

\section{New physics: the super heavy dark matter hypothesis} 
\label{sec:SHDM}

The extreme energies of UHECR, as high as $10^{11}$ GeV, eleven orders of magnitude above the proton mass and "only" eight below the Planck mass, are a unique workbench to probe new ideas, models and theories which show their effects at energies much larger than those ever obtained, or obtainable in the future, in accelerator experiments. An interesting class of exotic models that can be tested trough UHECR are top-down models for the production of these extremely energetic particles. The idea of top-down models arose in 90s to explain the lacking of the high energy suppression in the flux observed by the AGASA experiment (see \cite{Aloisio:2017ooo,Aloisio:2017qoo} and reference therein). Nowadays, with the firm experimental observation of the high energy suppression in the UHECR flux, the scientific case for super-heavy relics is connected with the problem of DM and cosmological observations. 

The leading paradigm to explain DM observations is based on the Weakly Interactive Massive Particle (WIMP) hypotheses, which consists of two basic assumptions: (i) WIMPs are stable particles of mass $M_\chi$ that interact weakly with the Standard Model (SM) particles; (ii) WIMPs are thermal relics, i.e. they were in Local Thermal Equilibrium (LTE) in the early Universe. Imposing that the WIMP density today is at the observed level, using a simple unitarity argument for the WIMP annihilation cross section $\sigma_{ann}\propto 1/M_{\chi}^2$, one obtains an estimate of the WIMP mass in the range of $10^{2}$ to  $10^{4}$ GeV. This result, also called the WIMP ``miracle", links the DM problem to the new physics scale expected in the context of the ``naturalness" argument for electroweak physics. Triggering in this way the strong hope that the search for WIMP DM may be connected to the discovery of new physics at the TeV scale. 
Searches for WIMP particles are ongoing through three different routes: direct detection, indirect detection, and accelerator searches. None of these efforts have discovered a clear WIMP candidate so far. In addition, no evidence for new physics has been observed at the Large Hadron Collider. Although not yet conclusive, the lack of evidence for WIMP may imply a different solution for the DM problem outside of the WIMP paradigm.

An alternative to WIMP models is represented by the scenario based on super-heavy relics produced due to time varying gravitational fields in the early universe: the so-called Super Heavy Dark Matter (SHDM). This approach is based on the possibility of particle production due to the non-adiabatic expansion of the background space-time acting on the vacuum quantum fluctuations (see \cite{Aloisio:2017qoo,Aloisio:2015lva} and references therein). In the framework of inflationary cosmologies, it was shown that particle creation is a common phenomenon, not tied to any specific cosmological scenario, that can play a crucial role in the solution to the DM problem as SHDM (labeled by $X$) can have $\Omega_X(t_0)\simeq 1$ \cite{Aloisio:2006yi,Kolb:2007vd}. This conclusion can be drawn under three general hypotheses: (i) SHDM in the early universe never reaches LTE; (ii) SHDM particles have mass of the order of the inflaton mass, $M_{\phi}$; and (iii) SHDM particles are long-living particles with a lifetime exceeding the age of the universe, $\tau_X\gg t_0$. 

Precision measurements of CMB polarisation and observations of UHECR up to energies $\simeq 10^{20}$ eV enable a direct experimental test of the three hypothesis listed above. As discussed in \cite{Aloisio:2015lva}, the first two hypothesis can be probed through the measurements of CMB polarisation. The third hypothesis, particle life-time longer than the age of the universe, is a general requirement of any DM model based on the existence of new particles. As in the case of WIMP, discrete gauge symmetries protecting particles from fast decays work equally well for SHDM particles (see \cite{Aloisio:2015lva,Aloisio:2006yi,Kolb:2007vd,Aloisio:2003xj} and references therein).

The best way to test the existence of SHDM is through the indirect detection of its annihilation and/or decay products (direct detection is unattainable). Since the annihilation cross section of a (point) particle is bounded by unitarity, $\sigma_{ann}\propto 1/M_X^2\sim 1/M_\phi^2$, the annihilation process results in an unobservable small annihilation rate \cite{Aloisio:2006yi}. 

If SHDM particles decay, under general assumptions on the underlying theory (see \cite{Aloisio:2015lva,Aloisio:2006yi,Kolb:2007vd,Aloisio:2003xj} and references therein), we can determine the composition and spectra of the standard model particles produced. Typical decay products are neutrinos, gamma rays and nucleons with a flat spectrum, that at the relevant energies can be approximated as $dN/dE \propto E^{-1.9}$, independently of the particle type, with a photon/nucleon ratio of about $\gamma/N\simeq 2\div 3$ and a neutrino nucleon ratio $\nu/N\simeq 3\div 4$, quite independent of the energy range \cite{Aloisio:2003xj}. The most constraining limits on SHDM lifetime are those coming from the (non) observation of UHE photons and, even to a lesser extent, neutrinos. Auger observations provide us with very stringent limits on the photons flux at energies above $10^{19}$ eV, which are at the level of $2\%$ \cite{Aglietta:2007yx}, this fact already constrains the SHDM life-time to values $\tau_X\ge 10^{21} \div 10^{22}$ yr depending on the underlying inflationary potential.

In the left panel of figure \ref{fig3}, as discussed in \cite{Aloisio:2015lva}, we plot the flux of UHECR coming from the decay of SHDM in a specific model of inflation with $M_{X}=4.5\times 10^{13}$ GeV and $\tau_X=2.2\times 10^{22}$ yr (solid lines); we also show the expected sensitivities of the proposed JEM-EUSO space mission (thick red solid line) \cite{Ebisuzaki:2014wka} and, for UHE neutrinos, of the upcoming ARA observatory (thick blue solid line) \cite{Allison:2014kha}. In the right panel we show the corresponding fraction over the total flux of protons, photons and neutrinos by the decay of SHDM. From figure \ref{fig3} it is clear that SHDM models can be effectively probed only by the next generation of UHECR experiments, those designed to maximise statistics at the highest energies \cite{Olinto:2017xbi}, together with new and more refined observations of the CMB polarisation pattern as they constrain the inflationary scenario.  

\section{Conclusions}
\label{sec:concl}

One of the main goals in the study of UHECR is a clear identification of the sources. In this ambit, as discussed above, a key observable is the mass composition. A pure proton composition is, theoretically, a natural possibility. Proton is the most abundant element in the universe and several different astrophysical objects, at present and past cosmological epochs, could provide efficient acceleration even if it requires very high luminosities and maximum acceleration energies. The complexity of the scenario based on a composition with heavy nuclei disfavours astrophysical sources placed at high redshift because of the lacking of heavy elements. To reproduce observations, mixed composition requires sources with flat injection spectra for heavy nuclei, or needs particular dynamics in the source environment, such as photo-disintegration on strong local photon fields. A remarkable feature of the mixed composition scenario is the relative low maximum energy required at the source.

After the first detection of astrophysical neutrinos at energies below ${\rm few}\times 10^{15}$ eV by the IceCube collaboration \cite{Aartsen:2015awa}, the study of HE and UHE neutrinos attracted a renewed interest. The observations of IceCube, being at relatively low energy, can be only marginally explained in the framework of cosmogenic neutrinos coming from UHECR interactions, also given the large uncertainties on the EBL background at high red-shift \cite{Aloisio:2015ega}. At high energies ($E\ge 10^{18}$) neutrino production is critically related with the mass composition of UHECR and with cosmological evolution of sources.

The study of UHECR has an impact also in fundamental physics, as it involves tests of models and theories that extend beyond the standard model of particle physics. This is the case of SHDM that, being a viable alternative to the WIMP paradigm for DM, can be tested only through the combined observations of the CMB polarisation pattern and UHECR at the highest energies \cite{Aloisio:2015lva}. 

We conclude highlighting the two principal avenues on which the study of UHECR should develop in the near future. From one hand, as discussed above, a firm experimental determination of the mass composition is an unavoidable step forward in this field of research. On the other hand, the highest energy regime, typically energies $E \ge 5\times 10^{19}$ eV, still remains less probed with not enough statistics to firmly detect possible anisotropies in the arrival directions and the exact shape of the suppression. Current technologies, such as Auger and (extended) TA, can reach, in several years, one order of magnitude more in the number of observed events at the highest energies, which seems not enough to firmly detect anisotropies, to probe new physics or to detect neutrinos. New technologies are needed and future space observatories, with improved photon detection techniques, like the proposal POEMMA (Probe of Extreme Multi-Messenger Astrophysics) \cite{Olinto:2017xbi}, promise a new era in the physics of UHECR, allowing the needed statistics to detect anisotropies, cosmogenic neutrinos and to probe new physics \cite{Aloisio:2017ooo,Aloisio:2017qoo,Olinto:2017xbi}.

\section*{References}
\bibliography{uhecr.bib}

\end{document}